\documentstyle[aps,prd,epsf]{revtex} 

\draft

\begin{document}
\twocolumn[\hsize\textwidth\columnwidth\hsize\csname
@twocolumnfalse\endcsname

\title{Massive warm dark matter}

\author{Steen Hannestad}

\address{NORDITA, Blegdamsvej 17, DK-2100 Copenhagen, Denmark}

\date{\today}

\maketitle

\begin{abstract}
Many independent high resolution simulations have indicated that the
standard collisionless cold dark matter model does not reproduce the
structure of observed present day galaxies well. Several possible
solutions in the form of modifications to the physics of the dark
matter particles have been proposed. One of the most promising is warm
dark matter (WDM), particles with significant thermal motion in the
early universe. It is usually assumed that such particles are
relativistically decoupled particles with a mass of approximately 1
keV. However, here we have investigated the possibility that much more
massive particles with highly non-thermal spectra could make up warm
dark matter. Several possible production mechanisms are reviewed and
the only one found to be viable is that the WDM is produced by the
non-relativistic decay of some massive species in the early
universe. Such very massive warm dark matter could possibly be
detected in direct detection experiments, as opposed to standard
thermal warm dark matter.
\end{abstract}

\vskip1.8pc]


\section{introduction}

The standard Big Bang model has been very successful in explaining the
large scale structure in the universe. An essential feature in the
model is dark matter. By far the most successful candidate for dark
matter is cold dark matter, collisionless particles which are so
massive that they are very non-relativistic during the entire
structure formation history \cite{peacock,gross}.

However, in the past few years high resolution N-body simulations have
shown that standard collisionless CDM apparently fails to reproduce
observations on galactic or smaller scales.  Observations indicate
that the halo of a galaxy like the milky way has almost an order of
magnitude fewer small satellite halos than is found in CDM simulations
\cite{moore,ghigna}.

Also, the central cores of dark matter halos have singular density
profiles, approaching almost $\rho \propto r^{-1.5}$ at small $r$
\cite{NFW96,FP94,N99,NS99,moore2}.  This is in contrast to
observations of low surface brightness galaxies, where the dark matter
halos apparently have almost constant density.

This seems like a very serious problem for the CDM structure formation
picture. However, at present it cannot be completely excluded that the
discrepancy is due to the low quality of either simulations or
observations. For instance very high resolution simulations which
include baryons have yet to be carried out \cite{simproblems}, and
rotation curve measurements of LSB galaxies may have greater
uncertainties than previously estimated \cite{obsproblems}.

However, the amount of observational data in conflict with the CDM
model is quite large and it seems a very real possibility that we may
have to abandon the simplest CDM collisionless models. The simplest
possibility (other possibilities, such as self-interacting CDM have
been proposed \cite{sidm}) is that power is suppressed on small
scales, either because of the initial power spectrum from inflation
\cite{KL99} or because of thermal motion of the dark matter particles.
The latter possibility corresponds to warm \cite{SS88,CSW96} or hot
dark matter and has received a lot of attention recently. WDM seems
able to explain many of the problems which CDM suffers from, notably
the substructure and the angular momentum problems
\cite{NSD,CAV,HD,SD99}.  However, it remains to be seen whether WDM
can also prevent singular cores from forming \cite{HD,HD2}.

Because WDM seems a quite promising candidate for dark matter it is
definitely worthwhile to look closer at the possible particle physics
mechanisms for producing it.  The simplest possibility is that the
warm dark matter is a relativistically decoupled species with a
thermal distribution function. This possibility has been reviewed a
number of times and we shall not discuss it further
\cite{SS88,CSW96,burns}. Rather we will investigate the intriguing
possibility that warm dark matter could have been produced
non-thermally. It is possible that dark matter particles with masses
much higher than thermal warm dark matter could have been produced
with enough thermal energy to make up warm dark matter. We shall
discuss three distinct possibilities for this, namely production by
decay of a heavy species \cite{HD}, production by annihilation of a
heavy species, and finally self heating by number changing self
interactions of the warm dark matter particles
\cite{CMH92,machacek,LSS,SD99}.  We then go on to discuss any possible
signatures that can distinguish massive warm dark matter from ordinary
warm dark matter.

Note that there is another possibility for non-thermal warm dark
matter, namely that the WDM particles are very light, but have been
produced with very low average momentum. This could happen if the WDM
particles are sterile neutrinos that have been produced via
oscillations with an active species \cite{fuller1} or by relativistic
decays of massive fermions into bosons \cite{madsen,hannestad1}.


\section{production mechanisms for massive warm dark matter}

A non-thermal warm dark matter candidate could have been produced in
several different ways via interactions with other particles in the
plasma. We here review three distinct possibilities:
self-interactions, decays and pair-annihilations. In the following,
$\phi$ is used to denote either the WDM field or the WDM particles,
irrespective of whether they are bosons or fermions.

\subsection{Production by self interactions}

Carlson {\it et al.} \cite{CMH92,machacek} suggested an interesting
way to heat cold dark matter relative to standard CDM, namely by
number changing self-interactions.  This was proposed as a means of
reconciling the $\Omega_m=1$ CDM model with large scale structure
observations. However, subsequently it was shown by de Laix, Scherrer
and Schaefer \cite{LSS} that this model produced a very poor fit to
observations.

At present the consensus is that the correct model for large scale
structure is a flat $\Lambda$CDM model \cite{peacock}, and so there is
no need to modify the physics of CDM to change the formation of large
scale structure.  Self-interacting dark matter could instead be
invoked as a possible means of explaining the observed small scale
structure. CDM with a large self-scattering cross section has been
proposed \cite{sidm}, and subsequently warm dark matter with
self-scatterings \cite{HS00}.  However, if the self-interactions are
sufficiently strong there are very likely number changing reactions as
well. This model is almost equivalent to that originally proposed by
Carlson {\it et al.} \cite{CMH92,machacek}.  If dark matter has such
strong self-interactions that number changing reactions occur in
equilibrium, the effective temperature of the species will drop much
slower than for other species.  In case of full thermal equilibrium
maintained by self-interactions, the distribution function is
(assuming Boltzmann statistics) $f = \exp(-E/T)$. Also, in thermal
equilibrium the total entropy within a comoving volume is
conserved. In this case, the effective temperature of the distribution
is
\begin{equation}
\frac{T}{m} = \left(3 \log(a) - K\right)^{-1},
\end{equation}
for $T \ll m$.  $a$ is the scale factor and $K$ is a constant.  This
shows that the effective temperature of the species only drops
logarithmically.  Note that this behaviour is in stark contrast with
that of the other species for which $T \propto a^{-1}$. Thus, the
particle distribution is heated to higher temperature than the
surrounding medium by transforming particles into kinetic energy.
This leads to a much larger free streaming length for such particles
than for thermally decoupled particles with the same mass.  The
observational bounds on the self-scattering cross section can be
translated into a rough bound on the number changing reactions,
depending on the specific type of WDM particle.

{\it Pseudoscalars} --- A likely candidate in this category is a
particle like the majoron (although there should be no coupling to
standard model fields) with a simple $\alpha \phi^4$ self-interaction
term. $4 \leftrightarrow 2$ number changing interactions take place
via a $\phi^6$ term in the effective Lagrangian \cite{CMH92}.  Based
on purely dimensional arguments the rate for this reaction should be
\cite{LSS}
\begin{equation}
\Gamma_{4 \leftrightarrow 2} \simeq n^3 \alpha^4/m^8,
\end{equation}
where $n$ denotes the number density.  In all reasonable scenarios we
should expect that the dimensionless coupling constant, $\alpha
\lesssim 1$.  However, we can also put another constraint on $\alpha$
from simple elastic scattering. The rate for elastic scattering of
these particles is
\begin{equation}
\Gamma \simeq n v \alpha^2/m^2,
\end{equation}
where $v$ is their relative velocity.  It has recently been shown in
numerical simulations that dark matter with a large elastic cross
section is not a good candidate because galaxies have very singular
halos and are too spherical \cite{sidm}.  A safe upper bound on the
2-body elastic scattering cross section is \cite{sidm}
\begin{equation}
\sigma \leq 10^{-22} m_{\rm GeV} \, {\rm cm}^2,
\end{equation} 
yielding an upper limit on $\alpha$
\begin{equation}
\alpha^2 \lesssim 2.6 \times 10^{-22} m_{\rm eV}^3.
\end{equation}

The quantity of interest here is the ratio of the number changing
interaction rate and the Hubble parameter, $\Gamma/H$. For
pseudo-scalars this can be written as
\begin{equation}
\frac{\Gamma}{H} \lesssim 1.1 \times 10^{30} (\Omega_\phi h^2)^3
\alpha^4 m_{\rm eV}^{-11} T_{\rm eV}^7 g_*^{-1/4},
\end{equation}
using the bound on $\alpha$.
$\Omega_\phi$ is the present-day density contribution of the WDM
and $g_*$ is the number of relativistic degrees of freedom present.
Freeze-out occurs at $\Gamma/H \simeq 1$, corresponding to the bound
\begin{equation}
m_{\rm eV}^{5/7} \lesssim 0.0017 (\Omega_\phi h^2)^{3/7} T_{\rm
freeze, {\rm eV}}.
\end{equation}
As soon as the number changing reactions freeze out the particles
become non-relativistic. Therefore, if this scenario is to work, we
must demand that $T_{\rm freeze} \lesssim 100-200$ eV, so that the WDM
has the correct amount of thermal motion. For pseudo-scalars this
translates into
\begin{equation}
m_{\rm eV} \lesssim 0.22 (\Omega_\phi h^2)^{3/5} \ll T_{\rm freeze,
{\rm eV}} .
\label{eq:mbound}
\end{equation}
The conclusion is that massive WDM cannot be produced in this fashion
unless special circumstances prevail. One way out is to tinker with
the $4 \leftrightarrow 2$ reaction rate, increasing it by a large
factor over the naive estimate. However, it seems safe to assume that
this type of heating is an unlikely production mechanism for WDM.

{\it Scalars} --- If the particles are scalars there could be a $3
\leftrightarrow 2$ number changing term of the form $\Gamma \simeq n^2
\alpha^3/m^5$ \cite{CMH92}.  In this case Eq.~(\ref{eq:mbound})
changes into

\begin{equation}
m_{\rm eV} \lesssim 4.1 (\Omega_\phi h^2)^{2/5} \\
\end{equation}
Again, the conclusion is that self-heating cannot be the production
mechanism.

{\it Fermions} --- For fermions the situation is quite different.  The
leading number changing reaction is $2 f + 2 \bar{f} \to f \bar{f}$.
The fermion-fermion interaction necessitates some new force carrier,
$\eta$.  If it is a boson with mass smaller than the fermion then the
reaction $f \bar{f} \to 2 \eta$ will dominate completely and the
fermion will quickly disappear by annihilation. Thus, the force
carrier would have to be more massive than $f$. But in that case the
reaction $2 f + 2 \bar{f} \to f \bar{f}$ will be suppressed by high
orders of $T/m_\eta \ll 1$. Thus we can expect the number changing
rate for fermions to be much smaller than for pseudoscalars and far
too small to be of any practical interest.

Thus, no matter what nature the dark matter particles have, it is very
unlikely that self-heating can increase the thermal energy
sufficiently that WDM results.

\subsection{Production by decay of a massive species}

Another, and perhaps more obvious way is to produce the WDM particles,
$\phi$, via decays of strongly non-relativistic particles, $X$
\cite{HD}.  We can write the present day thermal velocity of the warm
dark matter as
\begin{equation}
v = \frac{p}{m_\phi} \simeq \frac{T_{\gamma,0}}{m_\phi}
\left(\frac{g_{*,0}}{g_{*,d}}
\right)^{1/3}\left(\frac{m_X}{T_{\gamma}}\right)_{T=T_{\rm decay}}
\end{equation}
We can use this relation to find how much energy density was in the
$X$-particles when they decayed
\begin{equation}
\rho_{X,d} = \frac{m_X}{m_\phi} \rho_c \Omega_\phi
\left(\frac{g_{*,d}}{g_{*,0}}\right)\left(\frac{T_{\gamma,d}}{T_{\gamma,0}}
\right),
\end{equation}
where $\Omega_\phi$ is the present day contribution to $\Omega$ from
the warm dark matter.  From these relations we can calculate the ratio
$\rho_X/\rho_R$ at decay
\begin{equation}
\frac{\rho_X}{\rho_R} = 0.025 \Omega_\phi h^2 \left(\frac{v}{0.4 {\rm
km \, s}^{-1}}\right) \left(\frac{g_{*,d}}{g_{*,0}}\right)^{1/3}.
\end{equation}
From this we can conclude that the $X$-particles never dominated the
energy density of the universe prior to decay. However, the radiation
produced at decay adds a significant contribution to $\rho_R$ which
could be detectable.

If the decay took place prior to BBN ($z \sim 10^{12}$) the $\phi$
particles were highly relativistic during BBN and their
contribution to the radiation energy density is (in units of the
energy density of a standard massless neutrino species)
\begin{equation}
\Delta N_\nu = 0.185 \Omega_\phi h^2 \left(\frac{v}{0.4 {\rm km \,
s}^{-1}}\right).
\end{equation}
For $\Omega_\phi=0.3$ and $h=0.65$ this yields $\Delta N_\nu \simeq
0.023$, too small a perturbation to be detected. Even with high
precision measurements of the primordial abundances this value seems
out of reach \cite{lisi}.  If the decay is after BBN, the contribution
of $X$ during BBN is entirely negligible.

Next, one can ask whether this is detectable in the CMBR anisotropy.
At recombination the $\phi$-particles are already strongly
non-relativistic.  Their contribution to the radiation energy density
can be estimated as
\begin{equation}
\rho_{\phi,R} \simeq \frac{1}{2} v^2 \Omega_\phi h^2
\left(\frac{T_{\gamma}} {T_{\gamma,0}}\right)^3.
\end{equation}
Again, we can parameterize this, in units of neutrino species at
recombination, to be
\begin{equation}
\Delta N_\nu \simeq 1.7 \times 10^{-8} \left(\frac{v}{0.4 {\rm km \,
s}^{-1}}\right)^2 \left(\frac{T_{\gamma}}{T_{\gamma,0}}\right).
\end{equation}
At recombination this gives $\Delta N_\nu \simeq 2 \times 10^{-5}$,
which is too small to be detected, even with the upcoming
high-precision experiments MAP and PLANCK \cite{lopez}.

Note that thermally decoupled WDM gives the same contribution to the
cosmic radiation density as does the decay-produced WDM. Thus, this
effect would in no case allow us to distinguish the two different
types of dark matter (unless the decay is after BBN, in which case the
decay-produced WDM would not contribute to the radiation during BBN).

\subsection{Production by annihilation of a massive species}

The final production mechanism we shall discuss is via the
pair-annihilation of some massive fermion species. Here, the
$\phi$-particles are produced via $X \bar{X} \to 2 \phi$.  $\phi$ can
be either a boson or a fermion. The prime example would be a heavy
neutrino species annihilating into massless fermions.  Annihilations
will in general not produce $\phi$ with much greater than thermal
energies. The reason is that $\Gamma/H$ is always an increasing
function of $T$ for non-relativistic particles, so that very few high
energy annihilation products are produced (unless one invokes a
specific physical mechanism to prevent annihilations at high
temperature \cite{kkt}).

In any case, if warm dark matter is produced by the annihilation of a
massive species, the annihilation rate must be high enough that
practically all heavy particles disappear before the annihilation
freezes out. Otherwise the remaining heavy particles will quickly
begin to dominate the energy density, resulting in a standard CDM
scenario.  But in the limit of strong interactions, the specifics of
the annihilation reaction does not matter, the decay proceeds in
equilibrium.  The final energy density of the annihilation products
can then be found by solving the simple differential equations
(assuming $2 \leftrightarrow 2$ annihilation and Boltzmann statistics
of the involved species)
\begin{equation}
\frac{dn_{\rm TOT}}{dt} = \frac{dn_{\rm X}}{dt} + \frac{dn_{\rm
\phi}}{dt} = - 3 H n_{\rm TOT}
\end{equation}
\begin{equation}
\frac{d\rho_{\rm TOT}}{dt} = \frac{d\rho_{\rm X}}{dt} +
\frac{d\rho_{\rm \phi}}{dt} = - 3 H (\rho_{\rm TOT} + P_{\rm TOT}).
\end{equation}
\begin{equation}
f_{\rm X,\phi} = \exp(-(\sqrt{m_{\rm X,\phi}^2 + p^2}-\mu)/T),
\end{equation}
where $\mu$ and $T$ denote the common pseudo-chemical potential and
temperature of the two species.

Fig.~1 shows the outcome of the annihilation process. The average
energy of the light particles is slightly higher than the expected
$3T$, namely $\langle E_{\phi} \rangle = 3.78 T$.  However, this
clearly shows that any annihilation-product cannot act as strongly
non-thermal warm dark matter.

After reviewing these three possible production scenarios for
non-thermal WDM we conclude that the only plausible possibility is
that the WDM particles have been produced by the decay of some massive
species.

\begin{figure}[h]
\begin{center}
\epsfysize=7truecm\epsfbox{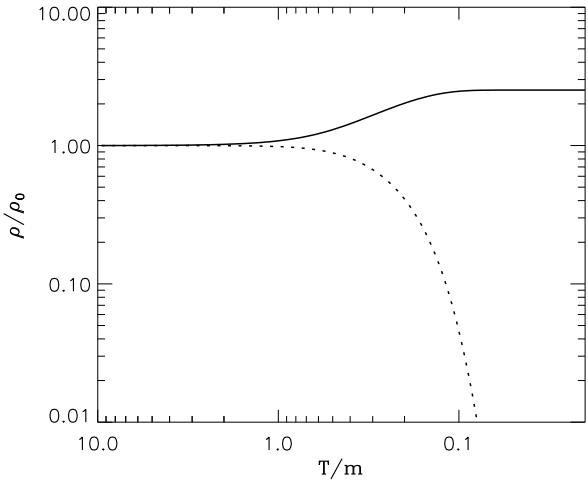}
\vspace{0truecm}
\end{center}
\caption{The evolution of the energy density of parent (dashed line)
and daughter (full line) for annihilation in equilibrium, parametrized
in units of the initial high temperature density for both species
(which is the same since they are in thermal equilibrium). Note that
the density is multiplied with $a^4$ so that it is constant for a
decoupled relativistic species.}
\label{fig1}
\end{figure}


\section{coupling to standard model fields}

Warm dark matter could in principle be coupled to standard model
fields. Thermal dark matter is much lighter than the $Z$-mass and
therefore any coupling to the standard model fields would likely show
up as a branching to $\phi$ in the $Z$-decay.  However, thermal WDM
cannot be detected in standard dark matter search experiments that
rely on nuclear recoil effects because the mass is much below
detection threshold \cite{gondolo}.  That would not need be the case
for massive warm dark matter candidates. In principle the WDM
particles could scatter on nuclei and be observed in direct detection
experiments.  However, the WDM should never come into equilibrium with
the standard model particles in the early universe. Otherwise a purely
thermal distribution results, and the end product is cold dark matter.
This puts a strict upper limit on the coupling between $\phi$ and
standard model fields.

As an example we calculate the energy equilibration rate for a Dirac
neutrino-like particle, $\phi$, in the early universe. The squared
matrix element for $\phi$ scattering on a massless standard model
fermion is roughly
\begin{equation}
\sum |M|^2 \simeq \eta G_F^2 (p_\phi \cdot p_f)(p_\phi^{'} \cdot
p_f^{'}),
\end{equation}
for energy transfers below $m_Z$. The dimensionless
parameter $\eta$ denotes the
effective coupling strength of $\phi$ to $Z$. From the above matrix
element we estimate the scattering rate for a $\phi$-particle to be
\begin{equation}
\Gamma_s = n_f \sigma \simeq \eta G_F^2 \langle E_\phi \rangle \langle
E_f \rangle T_f^3.
\end{equation}
The average energy of $\phi$ is much bigger than that of $f$. Thus, on
average the momentum transfer in each scattering is equivalent to
$\langle E_f \rangle$. The energy equilibration rate can then be
estimated as
\begin{equation}
\Gamma_E \simeq \eta G_F^2 \langle E_f \rangle^2 T_f^3 \simeq \eta
G_F^2 T_f^5.
\end{equation}
If the WDM particles are to stay out of thermal equilibrium then
\begin{equation}
\frac{\Gamma}{H} < 1
\end{equation}
must be fulfilled at all times after $\phi$ is produced.  This above
equation can be transformed into a bound on $\eta$
\begin{equation}
\eta < 2.8 g_*^{1/2} T_{\rm MeV}^{-3}.
\label{eq:etabound}
\end{equation}
Thus, the bound strengthens with increasing temperature and is
strongest at $T_d$, the decay temperature of the parent.

A direct detection experiment would typically use WIMP-nucleon elastic
scattering recoil. The elastic scattering cross section is given
approximately by \cite{gondolo}
\begin{equation}
\sigma \simeq \eta G_F^2 \mu^2,
\end{equation}
where $\mu = (m_N m_\phi)/(m_N + m_\phi)$ is the reduced mass of the
system.

The present generation of experiments have reached a limit of $\sigma
\simeq 1$ pb for masses of $\phi$ in the GeV range
\cite{DMexp}. In that case $\mu/m_N \simeq 1$ and the detection limit
corresponds roughly to
\begin{equation}
\eta \simeq 20.
\end{equation}
Using Eq.~(\ref{eq:etabound}) we find that
\begin{equation}
T_{\rm d,MeV} \lesssim 0.5
\end{equation}
if the $\phi$-particles were to have been detected in the present
experiments.  The next generation of dark matter experiments will
probably achieve about a factor $10^3$ better sensitivity,
yielding $T_{\rm d,MeV} \lesssim 5$.  In any case, if
this type of warm dark matter is to be detected in direct detection
experiments it must have been produced at the epoch of BBN or later,
otherwise the upper bound on the coupling to standard model fields
becomes so tight that no direct detection experiment will be able to
see it.


\section{discussion}

We have discussed various possible means of producing warm dark
matter, with specific focus on the possibility of very heavy warm dark
matter.  Dark matter self-heating was found to be excluded by present
observational bounds. Also, production by pair-annihilation of a
massive species cannot produce strongly non-thermal dark matter.  The
only viable production mechanism was found to be production by the
non-relativistic decay of a massive relic.

From a structure formation point of view there is very little
difference between massive warm dark matter and standard thermally
decoupled warm dark matter for the same average free streaming
length. Interestingly
,
as opposed to standard warm dark matter it could be possible to detect
massive warm dark matter in a direct detection experiment, because
the particle mass can easily be in the GeV range or above.
However, it was also found that such WDM would have to have been produced
at relatively low temperatures, otherwise the WDM distribution
would have been
brought into thermal equilibrium by interactions with the standard model
fields.

\acknowledgements{This work was supported by a grant from the
Carlsberg foundation}



\begin{references}
\bibitem{peacock}J.~A.~Peacock, ``Cosmological physics'', 
Cambridge University Press (1999).

\bibitem{gross}See for instance M.~Gross {\it et al.},
Mon.\ Not.\ R.\ Astron.\ Soc.\ {\bf 301}, 81 (1998). 

\bibitem{moore}B.~Moore {\it et al.},
Astrophys.\ J.\ Lett.\ {\bf 524}, 19 (1999).

\bibitem{ghigna}S.~Ghigna {\it et al.}, astro-ph/9910166 (1999).

\bibitem{NFW96}J.~F.~Navarro, C.~S.~Frenk and S.~D.~M.~White,
Astrophys.\ J.\ {\bf 462}, 563 (1996).

\bibitem{FP94}R.~Flores and J.~R.~Primack, Astrophys.\ J.\ {\bf 427},
L1 (1994).

\bibitem{N99}J.~F.~Navarro, astro-ph/9807084 (1998).

\bibitem{NS99}J.~F.~Navarro and M.~Steinmetz, astro-ph/9908114 (1999).

\bibitem{moore2}B.~Moore {\it et al.},
Mon.\ Not.\ R.\ Astron.\ Soc.\ {\bf 310}, 1147 (1999). 

\bibitem{simproblems}See for instance J.~Binney, O.~Gerhard and 
J.~Silk, astro-ph/0003199 (2000); J.~S.~Bullock, A.~V.~Kravtsov
and D.~H.~Weinberg, astro-ph/0002214 (2000) for a discussion of this
possibility.

\bibitem{obsproblems}See e.g.\ R.~A.~Swaters, B.~F.~Madore and 
M.~Trewhella, Astrophys.\ J.\ Lett.\ {\bf 531}, L107 (2000);
F.~C.~van den Bosch and R.~A.~Swaters, astro-ph/0006048 (2000).

\bibitem{sidm}D.~N.~Spergel and P.~J.~Steinhardt, 
astro-ph/9909386 (1999); S.~Hannestad, astro-ph/9912558;
A.~Burkert, Astrophys.\ J.\ Lett.\ {\bf 534}, 143 (2000); 
C.~Firmani {\it et al.}, Mon.\ Not.\ R.\ Astron.\ Soc.\ {\bf 315}, 29 (2000);
N.~Yoshida {\it et al.}, Astrophys.\ J.\ Lett.\ {\bf 535}, 103 (2000); 
B.~Moore {\it et al.}, Astrophys.\ J.\ Lett.\ {\bf 535}, 21 (2000);
J.~P.~Ostriker, astro-ph/9912548;
J.~Miralda-Escude, astro-ph/0002050;
B.~D.~Wandelt {\it et al.}, astro-ph/0006344;
R.~Dav{\'e} {\it et al.}, astro-ph/0006218;
N.~Yoshida {\it et al.}, astro-ph/0006134.

\bibitem{KL99}M.~Kamionkowski and A.~R.~Liddle,
astro-ph/9911103 (1999).

\bibitem{SS88}R.~Schaefer and J.~Silk,
Astrophys.\ J.\ {\bf 332}, 1 (1988).

\bibitem{CSW96}S.~Colombi, S.~Dodelson and L.~M.~Widrow,
Astrophys.\ J.\ {\bf 458}, 1 (1996).

\bibitem{NSD}V.~K.~Narayanan, D.~N.~Spergel and R.~Dav{\'e},
astro-ph/0005095.

\bibitem{CAV}P.~Colin, V.~Avila-Reese and O.~Valenzuela,
astro-ph/0004115.

\bibitem{HD}C.~J.~Hogan and J.~J.~Dalcanton, astro-ph/0002330.

\bibitem{SD99}J.~Sommer-Larsen and A.~Dolgov, astro-ph/9912166 (1999).

\bibitem{HD2}J.~J.~Dalcanton and C.~J.~Hogan, astro-ph/0004381.

\bibitem{burns}S.~D.~Burns, astro-ph/9711304.

\bibitem{CMH92}E.~D.~Carlson, M.~E.~Machanek and L.~J.~Hall,
Astrophys.\ J.\ {\bf 398}, 43 (1992).

\bibitem{machacek}M.~E.~Machacek, Astrophys.\ J.\ {\bf 431}, 41 (1994).

\bibitem{LSS}A.~A.~de Laix, R.~J.~Scherrer and R.~K.~Schaefer,
Astrophys.\ J.\ {\bf 452}, 495 (1995).

\bibitem{fuller1}X.~Shi and G.~M.~Fuller, 
Phys.\ Rev.\ Lett.\ {\bf 82}, 2832 (1999). 

\bibitem{madsen}J.~Madsen, Phys.\ Rev.\ Lett.\ {\bf 69}, 571 (1992). 

\bibitem{hannestad1}S.~Hannestad and J.~Madsen, 
Phys.\ Rev.\ D {\bf 55}, 4571 (1997).

\bibitem{HS00}S.~Hannestad and R.~J.~Scherrer, 
Phys.\ Rev.\ D {\bf 62}, 043522 (2000).

\bibitem{lisi}The present bound is about $\Delta N_\nu = 1$
(see for instance E.~Lisi, S.~Sarkar and F.~Villante,
Phys.\ Rev.\ D {\bf 59}, 123520 (1999).

\bibitem{lopez}R.~E.~Lopez {\it et al.}, Phys.\ Rev.\ Lett.\
{\bf 82}, 3952 (1999).

\bibitem{kkt}M.~Kaplinghat, L.~Knox and M.~S.~Turner, 
astro-ph/0005210 (2000).

\bibitem{gondolo}See for instance P.~Gondolo, astro-ph/9605290 (1996).

\bibitem{DMexp}See for instance the results presented from
DAMA (R.~Bernabei {\it et al.}, Phys.\ Lett.\ {\bf B480}, 23 (2000)),
CDMS (R.~Abusaidi {\it et al.}, Phys.\ Rev.\ Lett.\ {\bf 84}, 5699 (2000)
or CRESST (J.~Jochum {\it et al.}, hep-ex/0005003 (2000)).


\end{references}
\end{document}